\documentclass[aps,10pt,pra,twocolumn,showpacs,nofootinbib,superscriptaddress]{revtex4-1}
\usepackage{amsfonts}
\usepackage{amsmath}
\usepackage{amssymb}
\usepackage{graphicx}%
\usepackage{color}
\usepackage{cancel}
\usepackage{ulem}

\def\filetype{eps}
\providecommand{\U}[1]{\protect\rule{.1in}{.1in}}

\begin{document}
\title{Generalized Hartree-Fock-Bogoliubov Description of the Fr\"{o}hlich Polaron}
\author{Ben Kain}
\affiliation{Department of Physics, College of the Holy Cross, Worcester,
Massachusetts 01610 USA}
\author{Hong Y.\ Ling}
\affiliation{Department of Physics and Astronomy, Rowan University, Glassboro, New
Jersey 08028 USA}
\affiliation{Kavli Institute for Theoretical Physics, University of California, Santa
Barbara, California 93106 USA}
\affiliation{ITAMP, Harvard-Smithsonian Center for Astrophysics, Cambridge, Massachusetts 02138 USA}


\begin{abstract}
\noindent We adapt the generalized Hartree-Fock-Bogoliubov (HFB) method to an interacting many-phonon system free of impurities.  The many-phonon system is obtained from applying the Lee-Low-Pine (LLP) transformation to the Fr\"ohlich model which describes a mobile impurity coupled to noninteracting phonons.  We specialize our general HFB description of the Fr\"ohlich polaron to Bose polarons in quasi-1D cold atom mixtures. The LLP transformed many-phonon system distinguishes itself with an artificial phonon-phonon interaction which is very different from the usual two-body interaction.  We use the quasi-one-dimensional model, which is free of an ultraviolet divergence that exists in higher dimensions, to better understand how this unique interaction affects polaron states and how the density and pair correlations inherent to the HFB method conspire to create a polaron ground state with an energy in good agreement with and far closer to the prediction from Feynman's variational path integral approach than mean-field theory where HFB correlations are absent. 
\end{abstract}
\maketitle

\section{Introduction\newline}

Polarons emerge naturally from cold atom mixtures with an extreme population
imbalance where minority atoms are so outnumbered by majority atoms that they may be considered impurities submerged in a host medium. Polaron studies have undergone an exciting revival in recent years, sparked by
the experimental realization of polarons in mixtures of fermionic atoms
\cite{schirotzek09PhysRevLett.102.230402,nascimbene09PhysRevLett.103.170402},
with properties that are in excellent agreement with theoretical predictions
\cite{prokofev08PhysRevB.77.020408,mora09PhysRevA.80.033607}.  This
resurgence, which originally centered on the Fermi polaron problem where
background atoms are fermions (see
\cite{chevy10RepProgPhys.73.112401,massignan13arXiv:1309.0219} for a review), has spread rapidly to its bosonic cousin, where background atoms are
bosons, and led to recent detailed experimental studies \cite{jorgensen16arXiv:1604.07883,hu16arXiv:1605.00729}.  This so called Bose polaron problem has been the subject of theoretical studies using a variety of tools, including a weak coupling ansatz
\cite{Huang09ChinesePhysicsLetters.26.080302,shashi14PhysRevA.89.053617,kain14PhysRevA.89.023612}%
, a strong coupling approach
\cite{cucchietti06PhysRevLett.96.210401,sacha06PhysRevA.73.063604,casteels11LaserPhysics.21.1480}
involving the Landau and Pekar treatment
\cite{landau46ZhEkspTeorFiz.16.341,landau48ZhEkspTeorFiz.18.419},  a
variational approach \cite{tempere09PhysRevB.80.184504} based on Feynman's
path integral formalism \cite{feynman55PhysRev.97.660}, those
\cite{li14PhysRevA.90.013618,levinsen15PhysRevLett.115.125302} inspired by
a Chevy-type variational ansatz \cite{chevy06PhysRevA.74.063628}, exact
numerical simulation \cite{vlietinck15NewJournalOfPhysics.17.033023} based
upon the diagrammatic quantum Monte Carlo (MC) method
\cite{prokofev98PhysRevLett.81.2514,mishchenko00PhysRevB.62.6317} and the diffusion Monte Carlo method \cite{ardila_giorgini}, and a
systematic perturbation expansion
\cite{rath13PhysRevA.88.053632,sogaard15PhysRevLett.115.160401} involving
use of the $T$-matrix \cite{fetter71ManyParticleSystemsBook}.

Our interest here is with the Fr\"{o}hlich model
\cite{mahan00ManyParticlePhysicsBook}, a generic polaron model describing a
single mobile impurity interacting with a bath of bosonic particles.
 Interest in this model has remained virtually unabated ever since Landau and
Pekar \cite{landau46ZhEkspTeorFiz.16.341,landau48ZhEkspTeorFiz.18.419} likened
a polaron to an impurity dressed in a cloud of nearby phonons and Fr\"{o}hlich
\cite{frohlich54AdvPhys.3.325} formulated the problem in its present form more
than half a century ago (see \cite{alexandrov07Book} for a review). The recent upsurge of interest in the Bose polaron problem has once again brought the
Fr\"{o}hlich polaron to the forefront, examples of which include those in
Refs.\
\cite{cucchietti06PhysRevLett.96.210401,tempere09PhysRevB.80.184504,casteels11PhysRevA.83.033631}
for large (continuous) polarons and those in Refs.\
\cite{bruderer07PhysRevA.76.011605,bruderer08NewJournalOfPhysics.10.033015,privitera10PhysRevA.82.063614,tao15PhysRevA.92.063635}
for small (Holstein) polarons.

The present work has been motivated by recent studies
\cite{shashi14PhysRevA.89.053617,shchadilova14arXiv:1410.5691,grusdt15ScientificReports.5.12124,grusdt15arXiv:1510.04934}
that applied the well-known Lee, Low, and Pine (LLP) transformation
\cite{lee53PhysRev.90.297} to convert the Fr\"{o}hlich model, in which
impurities interact with non-interacting phonons, to the LLP-Fr\"ohlich  model (or LLP model for short), which describes an interacting phonon system free of impurity degrees of freedom.  When described within mean-field (MF) theory, the phonon ground state is a direct product of coherent
states at different momentum modes \cite{lee53PhysRev.90.297}---quantum
fluctuations (correlations), which can be of vital importance to a strongly
interacting system, are notably absent. We are particularly inspired by
recent attempts to overcome this weakness inherent in the MF product state by 
Shchadilova \textit{et al}.\ \cite{shchadilova14arXiv:1410.5691} using a
correlated Gaussian wave function (CGW) ansatz
\cite{altanhan93JPhysCondensMatter.5.6729,kandemir94JPhysCondensMatter.6.4505}
and Grusdt \textit{et al}.\
\cite{grusdt15ScientificReports.5.12124,grusdt15arXiv:1510.04934} using a
renormalization group (RG) approach \cite{hewson97HeavyFermionsBook}.

We adapt the self-consistent Hartree-Fock-Bogoliubov (HFB) approach to the interacting phonons in the LLP model. The HFB-based approach
shall be similar, in spirit, to the CGW ansatz, where various cross mode
correlations are automatically built in.  However, instead of independent
variables housed in a symmetric matrix, we parametrize quantum fluctuations
between various momentum modes with dependent variables (which will be the density and pair
correlation functions) housed in a single-particle density matrix. As a
result, instead of an unconstrained minimization we perform a constrained
minimization of energy with respect to the variational parameters
characterizing the quasiparticle vacuum defined via a
generalized Bogoliubov transformation. This approach allows Fr\"{o}hlich
polarons to be studied self consistently without having to introduce
additional small perturbative parameters. 

We test our HFB formalism by applying it to
Fr\"{o}hlich polarons in quasi-1D cold atom mixtures.  A remarkable feature of cold atom systems is that system parameters, such as
dimensionality and coupling strength, can be tuned precisely
\cite{bloch08RevModPhys.80.885}.  Potential avenues for realizing Bose polarons include Bose-Fermi mixtures with fermionic
impurities, e.g.\ $^{7}$Li-$^{6}$Li
\cite{schreck01PhysRevLett.87.080403,truscott01Science.291.2570,schreck01PhysRevLett.87.080403,ferrierBarbut14Science.345.1035}%
, $^{23}$Na-$^{6}$Li
\cite{hadzibabic02PhysRevLett.88.160401,stan04PhysRevLett.93.143001,
schuster12PhysRevA.85.042721}, $^{87}$Rb-$^{40}$K
\cite{ferrari02PhysRevLett.89.053202,roati02PhysRevLett.89.150403,inouye04PhysRevLett.93.183201,ferlaino06PhysRevA.73.040702}%
, $^{23}$Na-$^{40}$K \cite{park12PhysRevA.85.051602}, $^{87}$Rb-$^{6}$Li
\cite{silber05PhysRevLett.95.170408}, and $^{4}$He-$^{3}$He
\cite{macnamara06PhysRevLett.97.080404}, Bose-Bose mixtures with bosonic
impurities, e.g.\ $^{85}$Rb-$^{87}$Rb \cite{bloch01PhysRevA.64.021402}, $^{87}%
$Rb-$^{41}$K
\cite{modugno01Science.294.1320,catani08PhysRevA.77.011603,catani12PhysRevA.85.023623}%
, and $^{87}$Rb-$^{133}$Cs
\cite{mcCarron11PhysRevA.84.011603,spethmann12PhysRevLett.109.235301}, and
ion-Bose mixtures with ionic impurities, e.g.\ Ba$^{+}$-$^{87}$Rb
\cite{schmid10PhysRevLett.105.133202}.  1D systems have the nice property that
particle interactions can be resonantly enhanced by confinement-induced
resonance \cite{olshanii98PhysRevLett.81.938,bergeman03PhysRevLett.91.163201},
in addition to the usual Feshbach resonance \cite{chin10RevModPhys.82.1225}.
 Bose polarons in quasi-1D cold atoms have recently been experimentally
\cite{catani12PhysRevA.85.023623} and theoretically
\cite{catani12PhysRevA.85.023623,casteels12PhysRevA.86.043614} investigated.  The importance of HFB type quantum fluctuations in 1D Bose polarons has been stressed by Sacha and Timmermans \cite{sacha06PhysRevA.73.063604} in connection with impurity self-localization.

An important goal of the present work is to gain clean insight into how
phonon-phonon interactions in the LLP model affect quantum fluctuations, which in turn affect the underlying polaron states.  In 3D (as well as 2D \cite{casteels12PhysRevA.86.043614}) atomic models, computing the polaron energy
involves a momentum integral that contains an ultraviolet divergence
\cite{tempere09PhysRevB.80.184504}.  At the MF level
\cite{kain14PhysRevA.89.023612,shashi14PhysRevA.89.053617}, regularization
based on the Lippmann-Schwinger equation
\cite{fetter71ManyParticleSystemsBook} can remove this divergence, but such
regularization is unable to stem the log-divergence expected to arise in more
elaborate, e.g.\ RG and CGW, methods
\cite{grusdt15ScientificReports.5.12124,shchadilova14arXiv:1410.5691}.  By
contrast, 1D models do not suffer from such a problem.  Thus, testing the HFB
theory using 1D models provides us with a ``proof-of-principle" opportunity, allowing us to interpret our results in a manner free
of complications due to the ultraviolet divergence.

Our paper is organized as follows. In Sec.\ II we review and adapt the HFB
theory to the generic LLP model.  We construct the energy functional,
assuming the system to be in a generalized Bogoliubov quasiparticle vacuum
parameterized in terms of phonon fields describing a MF coherent state and
density and pair correlation functions describing quantum fluctuations.
 We apply the constrained Ritz variational principle to arrive at a set of
HFB equations specific to the LLP model.  In Sec.\ III we focus on a quasi-1D
Bose polaron in the context of cold atom physics and solve the problem using
our HFB theory self consistently.  For comparison we also solve the problem analytically using MF theory and numerically using Feynman's variational approach.  We discuss how phonon-phonon interactions
can enrich the polaron state and how quantum fluctuations
included in our HFB approach, which are absent in MF
theory, can help lower the polaron energy to a level in fairly good agreement
with Feynman's result, even in the regime of relatively light impurity and
strong coupling. We conclude in Sec.\ V.

\section{Theory: Self-Consistent HFB Formulation of Fr\"{o}hlich Polarons}

We begin with the generic Fr\"{o}hlich Hamiltonian
\cite{mahan00ManyParticlePhysicsBook}
\begin{equation}
\hat{H}^{\prime}=\frac{\mathbf{\hat{p}}^{2}}{2m_{I}}+\sum_{\mathbf{k}}%
\hbar\omega_{\mathbf{k}}\hat{b}_{\mathbf{k}}^{\dag}\hat{b}_{\mathbf{k}}%
+\sum_{\mathbf{k}}\frac{g_{\mathbf{k}}}{\sqrt{\mathcal{V}}}e^{i\mathbf{k\cdot
\hat{r}}}\left(  \hat{b}_{\mathbf{k}}+\hat{b}_{-\mathbf{k}}^{\dag}\right)
,\label{H-Frohlich}%
\end{equation}
which describes a single mobile impurity with mass $m_{I}$, momentum operator
$\mathbf{\hat{p}}$, and position operator $\mathbf{\hat{r}}$ interacting with
phonons with field operator $\hat{b}_{\mathbf{k}}$ for annihilating a phonon of
momentum $\hbar\mathbf{k}$ and energy $\hbar\omega_{\mathbf{k}}$, where
$g_{\mathbf{k}}$ is the impurity-phonon coupling strength and $\mathcal{V}$ is
the quantization length in 1D, area in 2D, and volume in 3D. 

Different systems are characterized with a different set of $\omega
_{\mathbf{k}}$ and $g_{\mathbf{k}}$.  In the solid state Einstein
model (containing longitudinal optical phonons), $\omega_{\mathbf{k}}$ is
modeled as a constant and $g_{\mathbf{k}}$ is modeled as inversely proportional to $k$.
 In the solid state acoustic model, $\omega_{\mathbf{k}}$ and $g_{\mathbf{k}%
}$ are approximated as proportional to $k$ and $\sqrt{k}$, respectively
\cite{peeters85PhysRevB.32.3515}.  In cold atom systems where impurities
are immersed in a Bose-Einstein condensate (BEC) of density $n_{B}$, phonons
are identified with Bogoliubov quasiparticles arising from BEC density
fluctuations, and $\omega_{\mathbf{k}}$ and $g_{\mathbf{k}}$ are given by
\begin{equation}
\omega_{\mathbf{k}}=v_{B}k\sqrt{1+\left(  \xi_{B}k\right)  ^{2}} \label{w_k}%
\end{equation}
and
\begin{equation}
g_{\mathbf{k}}=g_{IB}\sqrt{n_{B}\hbar k^{2}/\left(  2m_{B}\omega_{\mathbf{k}%
}\right)  },\label{g_k}%
\end{equation}
where $v_{B}=\sqrt{n_{B}g_{BB}/m_{B}}$ is the phonon speed, $\xi_{B}%
=\hbar/\sqrt{4m_{B}n_{B}g_{BB}}$ is the healing length, and $g_{BB}=4\pi
\hbar^{2}a_{BB}/m_{B}$ and $g_{IB}=4\pi\hbar^{2}a_{IB}/[  m_{IB}%
\equiv2m_{I}m_{B}/\left(  m_{I}+m_{B}\right)  ]  $ are, respectively,
boson-boson and bose-fermion interaction strengths with $a_{BB}$ and $a_{IB}$
$s$-wave scattering lengths.  In this cold atom case, Hamiltonian (\ref{H-Frohlich}) is measured
relative to the bare impurity-condensate interaction energy, $n_{B}g_{IB}$, which
accounts for the interaction of the impurity with the condensed bosons. In
reduced dimensions, $n_{B}$, $a_{BB}$, and $a_{IB}$ (hence $g_{BB}$ and
$g_{IB}$) are their effective versions for the corresponding dimensions. 

In what follows, we adopt a unit convention in which $\hbar=1$ (unless keeping $\hbar$ helps elucidate physics). 

\subsection{Fr\"{o}hlich Hamiltonian After Lee-Low-Pine Transformation}

The Lee-Low-Pine (LLP) transformation \cite{lee53PhysRev.90.297} is defined by
\begin{equation}
\mathcal{\hat{S}}=\exp\left(  i\sum_{\mathbf{k}}\mathbf{k}\hat{b}_{\mathbf{k}%
}^{\dag}\hat{b}_{\mathbf{k}}\mathbf{\cdot\hat{r}}\right) \label{S}
\end{equation}
and is a unitary transformation under which the phonon vacuum is invariant (since any power of  $\hat{b}_{\mathbf{k}}^{\dag}\hat{b}_{\mathbf{k}}$ gives
zero when acting on it).  Following \cite{shashi14PhysRevA.89.053617,shchadilova14arXiv:1410.5691,grusdt15ScientificReports.5.12124} we apply the LLP transformation to the Hamiltonian in (\ref{H-Frohlich}), $\hat{H}=\mathcal{\hat{S}}\hat{H}^{\prime}\mathcal{\hat{S}}^{-1}$, which gives
\begin{align}
\hat{H} &  =\frac{\left(  \mathbf{\hat{p}}-\sum_{\mathbf{k}}\mathbf{k}\hat
{b}_{\mathbf{k}}^{\dag}\hat{b}_{\mathbf{k}}\right)  ^{2}}{2m_{I}}\nonumber\\
&\qquad+  \sum_{\mathbf{k}}\omega_{\mathbf{k}}\hat{b}_{\mathbf{k}}^{\dag}\hat
{b}_{\mathbf{k}}+\frac{1}{\sqrt{\mathcal{V}}}\sum_{\mathbf{k}}g_{\mathbf{k}%
}\left(  \hat{b}_{\mathbf{k}}+\hat{b}_{-\mathbf{k}}^{\dag}\right)
,\label{H LLP}%
\end{align}
where we used $\mathcal{\hat{S}}\mathbf{\hat{p}}\mathcal{\hat{S}}%
^{-1}=\mathbf{\hat{p}}-\sum_{\mathbf{k}}\mathbf{k}\hat{b}_{\mathbf{k}}^{\dag
}\hat{b}_{\mathbf{k}}$ and $\mathcal{\hat{S}}\hat{b}_{\mathbf{k}%
}\mathcal{\hat{S}}^{-1}=\hat{b}_{\mathbf{k}}\exp(  -i\mathbf{k}\cdot\mathbf{\hat{r}} t)$.  Since $\hat{\mathbf{p}}$ commutes with $\hat{H}$ it is a constant of motion and may be replaced with its $c$-number equivalent $\mathbf{p}$, allowing Eq.\ (\ref{H LLP}) to be written as
\begin{align}
\hat{H} &  =\frac{p^{2}}{2m_{I}}+\frac{1}{\sqrt{\mathcal{V}}}\sum_{\mathbf{k}%
}g_{\mathbf{k}}\left(  \hat{b}_{\mathbf{k}}+\hat{b}_{-\mathbf{k}}^{\dag
}\right)  \nonumber\\
&\qquad+  \sum_{\mathbf{k}}\left(  \omega_{\mathbf{k}}+\frac{k^{2}}{2m_{I}}%
-\frac{\mathbf{k\cdot p}}{m_{I}}\right)  \hat{b}_{\mathbf{k}}^{\dag}\hat
{b}_{\mathbf{k}}+\hat{H}_{int},\label{H after LLP}%
\end{align}
where
\begin{equation}
\hat{H}_{int}=\frac{1}{2}\sum_{\mathbf{k},\mathbf{k}^{\prime}}\frac
{\mathbf{k}\cdot\mathbf{k}^{\prime}}{m_{I}}\hat{b}_{\mathbf{k}}^{\dag}\hat
{b}_{\mathbf{k}^{\prime}}^{\dag}\hat{b}_{\mathbf{k}^{\prime}}\hat
{b}_{\mathbf{k}}\label{four Boson interaction}%
\end{equation}
is a normal-ordered four-boson interaction term, representing the phonon-phonon interaction.

The Fr\"{o}hlich Hamiltonian (\ref{H-Frohlich}) prior to the LLP transformation
describes an impurity-phonon system where phonons are non-interacting but are
coupled to the impurity via terms involving $e^{i\mathbf{k}\cdot\mathbf{\hat{r}}}$, which account for the impurity recoil during emission and absorption
of a phonon.

The LLP transformation moves into a frame moving at a speed
determined by the total phonon momentum,
\begin{equation}
\mathbf{p}_{ph}=\sum_{\mathbf{k}}\mathbf{k}\left\langle \hat{b}_{\mathbf{k}%
}^{\dag}\hat{b}_{\mathbf{k}}\right\rangle .\label{p_ph}%
\end{equation}
This transformation is motivated by the fact that the total momentum (the
impurity momentum plus the total phonon momentum) is a constant of motion so
that in a moving frame defined by the total phonon momentum, the impurity
momentum $\mathbf{\hat{p}}$ becomes the total momentum and is thus a constant
of motion, replaceable with a $c$-number. 

As promised, the LLP transformation has transformed the Fr\"{o}hlich
Hamiltonian (\ref{H-Frohlich}) to Eq.\ (\ref{H after LLP}) which is free of
impurity degrees of freedom, but at the expense of phonons interacting via the
four-boson interaction in Eq.\ (\ref{four Boson interaction}).

\subsection{Generalized Bogoliubov Transformation and Polaron Energy
Functional}

From this point forward we describe our system as a many-body phonon system
free of impurities. The only indication of the impurity in the Hamiltonian
(\ref{H after LLP}) is $\mathbf{p}$, which we treat as a parameter (i.e.\
quantum number).  The impurity-phonon scattering term, $\sum_{\mathbf{k}}g_{\mathbf{k}}(  \hat{b}_{\mathbf{k}}+\hat{b}_{-\mathbf{k}}^{\dag
})$, being linear in the phonon field, leads to a nonzero average,
$\langle \hat{b}_{\mathbf{k}}\rangle \equiv z_{\mathbf{k}}$.
 It is convenient to move to the shifted phonon field, \
\begin{equation}
\hat{c}_{\mathbf{k}}=\hat{b}_{\mathbf{k}}-z_{\mathbf{k}}, \label{shifting}
\end{equation}
whose average vanishes, $\langle \hat{c}_{\mathbf{k}}\rangle =0$.
 The Hamiltonian (\ref{H after LLP}) then describes phonons in terms of
$z_{\mathbf{k}}$ and $\hat{c}_{\mathbf{k}}$.

In anticipation of the use of the Ritz variational principle in the next
subsection, we choose as the trial state, $\vert \phi\rangle $, 
the quasiparticle vacuum state defined by field operator $\hat
{d}_{\mathbf{k}}$, i.e.\ $\hat{d}_{\mathbf{k}}\vert \phi\rangle
=0$, where $\hat{d}_{\mathbf{k}}$ is defined through the generalized
Bogoliubov transformation, $\hat{d}_{\mathbf{k}}=\sum_{\mathbf{k}^{\prime}%
}(  U_{\mathbf{kk}^{\prime}}^{\ast}\hat{c}_{\mathbf{k}^{\prime}%
}-V_{\mathbf{kk}^{\prime}}^{\ast}\hat{c}_{\mathbf{k}^{\prime}}^{\dag})
$, which may equivalently be written%
\begin{equation}
\left(
\begin{array}
[c]{c}%
\hat{d}\\
\hat{d}^{\dag}%
\end{array}
\right)  =\mathcal{T}\left(
\begin{array}
[c]{c}%
\hat{c}\\
\hat{c}^{\dag}%
\end{array}
\right)  ,\label{Bogliubov T}%
\end{equation}
where%
\begin{equation}
\mathcal{T}=\left(
\begin{array}
[c]{cc}%
U^{\ast} & -V^{\ast}\\
-V & U
\end{array}
\right),  
\quad \mathcal{T}^{-1}=\left(
\begin{array}
[c]{cc}%
U^{T} & V^{\dag}\\
V^{T} & U^{\dag}%
\end{array}
\right)  .\text{ }\label{T and T inverse}%
\end{equation}
In Eqs.\ (\ref{Bogliubov T}) and (\ref{T and T inverse}), $\hat{c}$ ($\hat
{c}^{\dag}$) is a column vector with elements $\hat{c}_{i}\equiv\hat
{c}_{\mathbf{k}_{i}}$ ($\hat{c}_{i}^{\dag}\equiv\hat{c}_{\mathbf{k}_{i}}%
^{\dag}$) [a similar definition applies to $\hat{d}$ ($\hat{d}^{\dag}$)], and
$U$ ($V$) is a square matrix with matrix elements $U_{ij}\equiv U_{\mathbf{k}%
_{i}\mathbf{k}_{j}}$ ($V_{ij}\equiv V_{\mathbf{k}_{i}\mathbf{k}_{j}}$).  The
number of elements depends on the number of $\mathbf{k}$ values included in the calculation.

Defining%
\begin{equation}
\eta=\left(
\begin{array}
[c]{cc}%
I & 0\\
0 & -I
\end{array}
\right)  ,
\quad \gamma=\left(
\begin{array}
[c]{cc}%
0 & I\\
I & 0
\end{array}
\right)  ,\label{eta and gamma}%
\end{equation}
we note that for $\mathcal{T}$ in the form given in Eq.\
(\ref{T and T inverse}), $\gamma\mathcal{T}\gamma=\mathcal{T}^{\ast}$
holds automatically and the only requirement for the Bogoliubov transformation
(\ref{Bogliubov T}) to remain canonical is
\begin{equation}
\mathcal{T}\eta\mathcal{T}^{\dag}\eta=1,\label{cannonical condition}%
\end{equation}
which, together with Eq.\ (\ref{T and T inverse}), amounts to requiring $U$ and
$V$ to obey
\begin{subequations}
\label{UV}%
\begin{align}
UU^{\dag}-VV^{\dag} &  =I, & UV^{T}-VU^{T}&=0,\label{UV 1}\\
U^{\dag}U-V^{T}V^{\ast} &  =I, & U^{T}V^{\ast}-V^{\dag}U&=0.\label{UV2}%
\end{align}

Let $z_{i}\equiv z_{\mathbf{k}_{i}}$, $\omega_{i}\equiv\omega_{\mathbf{k}_{i}%
}$, $g_{i}\equiv g_{\mathbf{k}_{i}}$, and $\sum_{i}\equiv$ $\sum
_{\mathbf{k}_{i}}$.  The average of the Hamiltonian in the quasiparticle
vacuum, $E_{p}\equiv\langle \phi\vert \hat{H}\vert
\phi\rangle $, then reads
\end{subequations}
\begin{widetext}
\begin{align}
E_{p} &  =\frac{p^{2}}{2m_{I}}+\sum_{i}\left(  \omega_{i}-\frac{\mathbf{k}%
_{i}\mathbf{\cdot p}}{m_{I}}+\frac{k_{i}^{2}}{2m_{I}}\right)  \left\vert
z_{i}\right\vert ^{2}+\sum_{i}\frac{g_{i}}{\sqrt{\mathcal{V}}}\left(
z_{i}+z_{i}^{\ast}\right)  +\sum_{i,j}\frac{\mathbf{k}_{i}\cdot\mathbf{k}_{j}%
}{2m_{I}}\left\vert z_{i}\right\vert ^{2}\left\vert z_{j}\right\vert
^{2}\nonumber\\
&\qquad+  \sum_{i}\left(  \omega_{i}-\frac{\mathbf{k}_{i}\mathbf{\cdot p}}{m_{I}%
}+\frac{k_{i}^{2}}{2m_{I}}+\frac{\mathbf{k}_{i}}{m_{I}}\cdot\sum_{j}
\mathbf{k}_j
\left\vert
z_{j}\right\vert ^{2}\right)  \rho_{ii}+\sum_{ij}\frac{\mathbf{k}_{i}%
\cdot\mathbf{k}_{j}}{2m_{I}}\left(  z_{i}^{\ast}z_{j}^{\ast}\kappa_{ij}%
+z_{i}z_{j}\kappa_{ij}^{\ast}+z_{i}^{\ast}z_{j}\rho_{ij}+z_{i}z_{j}^{\ast}%
\rho_{ij}^{\ast}\right)  \nonumber\\
&\qquad+  \sum_{ij}\frac{\mathbf{k}_{i}\cdot\mathbf{k}_{j}}{2m_{I}}\left(
\kappa_{ij}^{\ast}\kappa_{ij}+\rho_{ij}^{\ast}\rho_{ij}+\rho_{ii}\rho
_{jj}\right)  ,\label{<H>}%
\end{align}
\end{widetext}
where we have introduced the single-particle density matrix $\rho$ and
single-particle pair matrix $\kappa$ of state $\left\vert \phi\right\rangle $
whose matrix elements are defined, respectively, as
\begin{subequations}
\label{rho and kappa}%
\begin{align}
\rho_{ij} &  =\rho_{ij}^{\dag}=\left\langle \phi\right\vert \hat{c}_{j}^{\dag
}\hat{c}_{i}\left\vert \phi\right\rangle ,\\
\kappa_{ij} &  =\kappa_{ij}^{T}=\left\langle \phi\right\vert \hat{c}_{j}%
\hat{c}_{i}\left\vert \phi\right\rangle ,
\end{align}
which become, with the help of the generalized Bogoliubov transformation
(\ref{Bogliubov T}),
\end{subequations}
\begin{subequations}
\label{rho and kappa 1}%
\begin{align}
\rho_{ij} &  =\left(  V^{\dag}V\right)  _{ij}=\sum_{n}V_{ni}^{\ast}V_{nj},\\
\kappa_{ij} &  =\left(  V^{\dag}U\right)  _{ij}=\sum_{n}V_{ni}^{\ast}U_{nj}.
\end{align}
The first line in Eq.\ (\ref{<H>}) follows from the part of the Hamiltonian
that is independent of the field operators ($\hat{c}_\mathbf{k},\hat
{c}_\mathbf{k}^\dag$), the second line follows from the part quadratic in field
operators ($\hat{c}_\mathbf{k},\hat{c}_\mathbf{k}^\dag$), and the last
line represents the average of the four-boson term, $\hat{c}_{\mathbf{k}%
}^{\dag}\hat{c}_{\mathbf{k}^{\prime}}^{\dag}\hat{c}_{\mathbf{k}^{\prime}}%
\hat{c}_{\mathbf{k}}$, which can be computed using Wick's theorem \cite{fetter71ManyParticleSystemsBook}.

\subsection{Ritz Variational Principle and Self-Consistent HFB Equations}

The self-consistent HFB method is typically employed to solve many-body
problems with a fixed (average) number of particles in nuclear and condensed
matter physics \cite{ring04Book,blaizot96QuantumTheoryBook}. In
comparison, the average number of phonons in our system is not given a priori;
it depends on the impurity and phonon interaction and is therefore unknown and
must be determined self consistently.  We thus use a canonical, instead of
grand canonical, Hamiltonian, which explains the absence of a chemical
potential in the energy functional (\ref{<H>}) compared to the usual HFB formulation.

Minimizing the energy in Eq.\ (\ref{<H>}) with respect to $\left(
z_{\mathbf{k}}\text{,}z_{\mathbf{k}}^{\ast}\right)  $, we arrive at a matrix
equation,
\end{subequations}
\begin{equation}
\left(
\begin{array}
[c]{cc}%
C & D\\
D^{\ast} & C^{\ast}%
\end{array}
\right)  \left(
\begin{array}
[c]{c}%
z\\
z^{\ast}%
\end{array}
\right)  =-\frac{1}{\sqrt{\mathcal{V}}}\left(
\begin{array}
[c]{c}%
g\\
g
\end{array}
\right)  ,\label{CD equation}%
\end{equation}
where $C=C^{\dag}$ and $D=D^{T}$ are matrices defined as
\begin{subequations}
\label{C and D}%
\begin{align}
C_{ij} &  =\mathcal{C}_{i}\delta_{i,j}+\frac{\mathbf{k}_{i}\cdot\mathbf{k}%
_{j}}{m_{I}}\rho_{ij},\\
D_{ij} &  =\frac{\mathbf{k}_{i}\cdot\mathbf{k}_{j}}{m_{I}}\kappa_{ij}.
\end{align}
Here,
\end{subequations}
\begin{equation}
\mathcal{C}_{i}=\omega_{i}-\frac{\mathbf{k}_{i}\cdot\left(  \mathbf{p}%
-\mathbf{p}_{ph}\right)  }{m_{I}}+\frac{k_{i}^{2}}{2m_{I}},\label{H_i}%
\end{equation}
which is the only surviving term in the MF theory when density and pair
correlation functions $\rho$ and $\kappa$ are neglected, and
\begin{equation}
\mathbf{p}_{ph}=\sum_{j}\mathbf{k}_{j}\left(  \left\vert z_{j}\right\vert
^{2}+\rho_{jj}\right)  \label{p_ph 1}%
\end{equation}
is the expectation value of the total phonon momentum [Eq.\ (\ref{p_ph})]
with respect to the quasiparticle vacuum $\left\vert \phi\right\rangle $.

The next step would normally be to minimize the energy with respect to $\rho$
and $\kappa$, but a word of caution is in order---$\rho$ and $\kappa$ cannot
be treated as independent variables.  This is because $\rho$ and $\kappa$ are made up of $U$ and $V$ [Eq.\ (\ref{rho and kappa 1})], which are not
independent [Eq.\ (\ref{UV})]. This may be contrasted with the correlated
Gaussian wave function approach \cite{shchadilova14arXiv:1410.5691} where the energy functional is parameterized in terms of a symmetric matrix with
independent parameters.  The restrictions imposed on $\rho$ and
$\mathbf{\kappa}$ can be understood, perhaps most conveniently, with the help
of the generalized density matrix \cite{blaizot96QuantumTheoryBook}
\begin{equation}
\mathcal{R}=\left\langle \phi\right\vert \left(
\begin{array}
[c]{cc}%
\hat{c}_{j}^{\dag}\hat{c}_{i} & \hat{c}_{j}\hat{c}_{i}\\
\hat{c}_{j}^{\dag}\hat{c}_{i}^{\dag} & \hat{c}_{j}\hat{c}_{i}^{\dag}%
\end{array}
\right)  \left\vert \phi\right\rangle =\left(
\begin{array}
[c]{cc}%
\rho & \kappa\\
\kappa^{\ast} & 1+\rho^{\ast}%
\end{array}
\right)  , \label{R definition}%
\end{equation}
for field $\hat{c}_{\mathbf{k}}$, and
\begin{equation}
\mathcal{R}^{\prime}=\left\langle \phi\right\vert \left(
\begin{array}
[c]{cc}%
\hat{d}_{j}^{\dag}\hat{d}_{i} & \hat{d}_{j}\hat{d}_{i}\\
\hat{d}_{j}^{\dag}\hat{d}_{i}^{\dag} & \hat{d}_{j}\hat{d}_{i}^{\dag}%
\end{array}
\right)  \left\vert \phi\right\rangle =\left(
\begin{array}
[c]{cc}%
0 & 0\\
0 & 1
\end{array}
\right)  ,
\end{equation}
for quasiparticle field $\hat{d}_{\mathbf{k}}$.  By virtue of the Bogoliubov
transformation in Eq.\ (\ref{Bogliubov T}), $\mathcal{R}^{\prime}$ is linked to
$\mathcal{R}$ according to
\begin{equation}
\mathcal{R}^{\prime}=\mathcal{TRT}^{\dag},
\end{equation}
which, together with Eq.\ (\ref{cannonical condition}), means that
\begin{equation}
\left(  \eta\mathcal{R}\right)  ^{2}=-\eta\mathcal{R}. \label{constraint}%
\end{equation}
Equation (\ref{constraint}) encapsulates all relations among $\rho\, $and
$\kappa$.

We now minimize the total energy $E_{p}$ in Eq.\ (\ref{<H>}) with respect to
$\mathcal{R}$ (or equivalently $\rho$ and $\kappa$) subject to condition
(\ref{constraint}), i.e.,%
\begin{equation}
\delta\left\{  E_{p}-\text{Tr}\left[  \Lambda\left(  \left(  \eta
\mathcal{R}\right)  ^{2}+\eta\mathcal{R}\right)  \right]  \right\}  =0,
\end{equation}
where $\Lambda$ is a matrix of Lagrangian multipliers implementing constraint (\ref{constraint}).  By carrying out the variation explicitly and
then eliminating $\Lambda$, we arrive at the HFB equation%
\begin{equation}
\left[  \eta\mathcal{M},\mathcal{R}\eta\right]  =0, \label{HFB equation}%
\end{equation}
where
\begin{equation}
\mathcal{M}=\left(
\begin{array}
[c]{cc}%
A & B\\
B^{\ast} & A^{\ast}%
\end{array}
\right)  ,
\end{equation}
and $A=A^{\dag}$ and $B=B^{T}$ are matrices defined below in Eq.\
(\ref{A and B}). 

At this point, we observe that the matrices in Eqs.\ (\ref{C and D}) and (\ref{A and B}) are local in the sense that a matrix element in the $i$th row
and $j$th column is determined by the matrix elements of $\rho$ and $\kappa$ in the same row
and same column, e.g., $C_{ij}$ depends on $\rho_{ij}$ but not on $\rho_{i^{\prime}\neq i,j^{\prime}\neq j}$.  This local property is unique to the four-boson term in Eq.\
(\ref{four Boson interaction}) as we now explain.  If particles were to
interact via, for example, the usual two-body $s$-wave potential, the four-boson
term would be in the form $\sum_{\mathbf{k},\mathbf{k}^{\prime},\mathbf{q}%
}\hat{b}_{\mathbf{k}+\mathbf{q}}^{\dag}\hat{b}_{\mathbf{k}^{\prime}%
-\mathbf{q}}^{\dag}\hat{b}_{\mathbf{k}^{\prime}}\hat{b}_{\mathbf{k}}$, in
which momentum $\mathbf{q}$ is exchanged in each scattering event.  The
local property would then not hold, e.g., $C_{ij}$ would depend not only on
$\rho_{ij}$ but also on $\rho_{i^{\prime}\neq i,j^{\prime}\neq j}$.  The
four-boson term in Eq.\ (\ref{four Boson interaction}) is, however, of a very
different origin, arising artificially from the LLP transformation:\ in the ``boosted" LLP frame, phonons appear to interact without momentum exchange, i.e.\ $\mathbf{q}=0$.  It is this lack of momentum exchange that is responsible for the local property, and it allows us to formulate a much simplified HFB description of the Fr\"{o}hlich model compared to if the model had the usual two-body interaction.

Returning to Eq.\ (\ref{HFB equation}), the fact that $\eta\mathcal{M}$ and
$\mathcal{R}\eta$ commute means that solving for $\mathcal{R}$ from Eq.\
(\ref{HFB equation}) amounts to finding a set of simultaneous eigenstates of
$\eta\mathcal{M}$ and $\mathcal{R}\eta$.  Consider first the eigenstates of
$\eta\mathcal{M}$, which, because $\gamma\mathcal{M}\gamma=\mathcal{M}^{\ast}%
$, are grouped into pairs with eigenvalues $\pm w_{n}$; for each eigenstate
$\left\vert w_{n}^{+}\right\rangle $ with a positive (real)
eigenvalue, $w_{n}>0$, there exists an eigenstate $\left\vert w_{n}
^{-}\right\rangle =\gamma\left\vert w_{n}^{+}\right\rangle ^{\ast}$ with the negative of that eigenvalue, $-w_{n}$:
\begin{equation}
\eta\mathcal{M}\left\vert w_{n}^{+}\right\rangle =w_{n}\left\vert w_{n}%
^{+}\right\rangle,
\quad
\eta\mathcal{M}\left\vert w_{n}^{-}\right\rangle
=-w_{n}\left\vert w_{n}^{-}\right\rangle . \label{w+ w-}%
\end{equation}
The set of states $\left\vert w_{n}^{\pm}\right\rangle $ is complete in the
sense that they obey orthonormality conditions with metric $\eta$:%
\begin{equation}
\left\langle w_{n}^{\pm}\right\vert \eta\left\vert w_{m}^{\pm}\right\rangle
=\pm\delta_{n,m},
\quad
\left\langle w_{n}^{+}\right\vert \eta\left\vert w_{m}%
^{-}\right\rangle =0.
\end{equation}

Next consider the eigenstates of $\mathcal{R}\eta$.  From Eq.\ (\ref{constraint}) we have $\left(  \mathcal{R}\eta\right)  ^{2}%
=-\mathcal{R}\eta$, which allows us to divide the eigenstates into two
groups, one group with eigenvalue $0$ and the other group with eigenvalue
$-1$. $\mathcal{R}$, the solution to the HFB equation (\ref{HFB equation}),
must then take the form
\begin{equation} \label{R self-consistent}
\mathcal{R}=\sum_{n}\left\vert w_{n}^{-}\right\rangle \left\langle w_{n}%
^{-}\right\vert =\sum_{n}\gamma\left\vert w_{n}^{+}\right\rangle ^{\ast
}\left\langle w_{n}^{+}\right\vert ^{\ast}\gamma
\end{equation}
in the space spanned by $\left\{  \left\vert w_{n}^{\pm}\right\rangle
\right\}  $, from which we easily find that $\left\vert w_{n}^{\pm
}\right\rangle $ are also eigenstates of $\mathcal{R}\eta$:
\begin{equation}
\mathcal{R}\eta\left\vert w_{n}^{+}\right\rangle =0\left\vert w_{n}%
^{+}\right\rangle ,
\quad
\mathcal{R}\eta\left\vert w_{n}^{-}\right\rangle
=-1\left\vert w_{n}^{-}\right\rangle .
\end{equation}

We have now defined two expressions for $\mathcal{R}$, one in Eq.\
(\ref{R self-consistent}) in terms of the eigenstates of $\eta\mathcal{M}$
[Eq.\ (\ref{w+ w-})] and the other earlier in Eq.\ (\ref{R definition}) in terms
of the $U$ and $V$ matrices [Eq.\ (\ref{rho and kappa 1})]. Self consistency
requires that they be equivalent, which can be accomplished by making the
$n$th row of matrices $U$ and $V$ in Eq.\ (\ref{T and T inverse}) equal to the
$n$th eigenstate $\vert w_{n}^{+}\rangle =(  U_{n}%
,V_{n})  ^{T}$ of Eq.\ (\ref{w+ w-}) or by explicitly constructing $U$
and $V$ from those states with positive eigenvalues in Eq.\ (\ref{w+ w-}), which in matrix form is
\begin{equation}
\left(
\begin{array}
[c]{cc}%
A & B\\
-B^{\ast} & -A^{\ast}%
\end{array}
\right)  \left(
\begin{array}
[c]{c}%
U\\
V
\end{array}
\right)  =w\left(
\begin{array}
[c]{c}%
U\\
V
\end{array}
\right)  ,\label{AB equation}%
\end{equation}
where $A=A^{\dag}$ and $B=B^{T}$ are defined as
\begin{subequations}
\label{A and B}%
\begin{align}
A_{ij} &  =\mathcal{C}_{i}\delta_{i,j}+\frac{\mathbf{k}_{i}\cdot\mathbf{k}%
_{j}}{m_{I}}\left(  \rho_{ij}+z_{i}z_{j}^{\ast}\right)  ,\\
B_{ij} &  =\frac{\mathbf{k}_{i}\cdot\mathbf{k}_{j}}{m_{I}}\left(  z_{i}%
z_{j}+\kappa_{ij}\right)  ,
\end{align}
with ($\rho_{ij},\kappa_{ij}$) and $\mathcal{C}_{i}$ already given in Eqs.\
(\ref{rho and kappa 1}) and (\ref{H_i}), respectively.

In summary, following the generalized HFB approach
\cite{ring04Book,blaizot96QuantumTheoryBook}, we have arrived at the closed
set of equations (\ref{rho and kappa 1}), (\ref{CD equation}), (\ref{p_ph 1}),
and (\ref{AB equation}), which constitutes our HFB formulation of Fr\"{o}hlich
polarons.  Although we apply these equations to cold atom systems in the next section, we stress that they were derived generally, and we have in mind their widespread use for the many applications of the Fr\"ohlich model.

\section{Application: Quasi-1D Bose Polarons}

This section is devoted to the study of a Bose polaron in a 1D cold atom
mixture where atoms are confined, by sufficiently high harmonic trap
potentials along the transverse dimensions, to a 1D waveguide where the
transverse degrees of freedom are ``frozen" to the zero-point oscillation.  This problem has been
investigated by Casteels \textit{et.\ al.}\ \cite{casteels12PhysRevA.86.043614} at finite temperature using
Feynman's variational method \cite{feynman55PhysRev.97.660}.  In the present work, we focus exclusively on the zero temperature limit.

We will explore various polaron properties in terms of the
polaronic coupling constant $\alpha^{\left(  1\right)  }$ [defined below in
Eq.\ (\ref{alpha})] and the boson-impurity mass ratio $m_{B}/m_{I}$.  The
former can be tuned via a combination of Feshbach resonance and confinement-induced resonance \cite{olshanii98PhysRevLett.81.938,bergeman03PhysRevLett.91.163201}
while the latter can be treated practically as a tunable parameter owing to
the rich existence of atomic elements and their isotopes in nature.  The Fr\"ohlich Hamiltonian omits a quartic interaction term (which is quadratic in both the impurity and the BEC operators).  This term describes scattering between the impurity and a Bogolubov mode and is essential to correctly describe strong interactions near a Feshbach resonance between the impurity and BEC.  The absence of this term places an upper bound on the impurity-BEC coupling strength.  A thorough analysis of 3D Bose polarons in cold atomic systems [37, 38] indicates that an intermediate coupling regime is accessible to current technology involving interspecies Feshbach resonance. As in other studies of strongly interacting Bose polarons (see e.g.\ \cite{cucchietti06PhysRevLett.96.210401, tempere09PhysRevB.80.184504, casteels11PhysRevA.83.033631, shchadilova14arXiv:1410.5691, grusdt15ScientificReports.5.12124, grusdt15arXiv:1510.04934}), we extend our theory into the strongly interacting regime with the understanding that such results have only qualitative meaning.

\subsection{Polaron States}

Before presenting the full HFB description, we first consider the MF
description of polarons, which is described by $z_{k}$, governed by Eq.\
(\ref{CD equation}), in the MF limit where all correlations vanish (i.e., $\kappa=$ $\rho=0$):
\end{subequations}
\begin{equation}
z_k=-\frac{1}{\sqrt{\mathcal{V}}}\frac{g_k}{\omega_k+\frac{k^{2}}%
{2m_{I}}-\frac{k\left(  p-p_{ph}\right)  }{m_{I}}}, \label{MF z}%
\end{equation}
where $k$ ranges from $-\infty$ to $+\infty$.  The only unknown in Eq.\
(\ref{MF z}) is $p_{ph}$, which is given by Eq.\ (\ref{p_ph 1}).  If we can
solve for $p_{ph}$, $z_{k}$ is completely determined.  Inserting Eq.\ (\ref{MF z}) into
Eq.\ (\ref{p_ph 1}) and moving to an integral in terms of the scaled quantities
$\left(  \bar{k},\bar{p},\bar{p}_{ph}\right)  =\left(  k,p,p_{ph}\right)
\xi_{B}$ and $\bar{m}_{B}=m_{B}/m_{I}$, we obtain%

\begin{align}
\bar{p}_{ph} &  =4\alpha^{\left(  1\right)  }\bar{m}_{B}\left(  1+\bar{m}%
_{B}\right)  ^{2}\left(  \bar{p}-\bar{p}_{ph}\right)  \int_{0}^{\infty}%
d\bar{k}\nonumber\\
&\times  \frac{1+\bar{m}_{B}\bar{k}/\sqrt{1+\bar{k}^{2}}}{\left[  \left(
\sqrt{1+\bar{k}^{2}}+\bar{m}_{B}\bar{k}\right)  ^{2}-4\bar{m}_{B}^{2}\left(
\bar{p}-\bar{p}_{ph}\right)  ^{2}\right]  ^{2}},\label{p_ph bar}%
\end{align}
where
\begin{equation}
\alpha^{\left(  1\right)  }=a_{IB}^{2}\xi_{B}/a_{BB}\label{alpha}%
\end{equation}
is the 1D dimensionless polaron coupling constant.\footnote{The 1D coupling constant used by Casteels \textit{et.\ al.}\ in \cite{casteels12PhysRevA.86.043614} (also labeled $\alpha^{(1)}$) equals $2\sqrt{2}\pi\alpha^{(1)}$.}  Evaluating the integral (\ref{p_ph bar}), we find that $p_{ph}$ corresponds to the root of the following transcendental equation:
\begin{widetext}
\begin{equation}
\bar{p}_{ph}=\alpha^{\left(  1\right)  }\left(  1+\bar{m}_{B}\right)
^{2}  \left\{
\begin{array}{ll}
\frac{\bar{b}\bar{m}_{B}}{\left(  1-\bar{b}^{2}\right)  \bar{r}}+\frac{\bar
{b}}{\bar{r}\sqrt{\left\vert \bar{r}\right\vert }}\left(  \tanh^{-1}%
\frac{1+\bar{m}_{B}+\bar{b}}{\sqrt{\left\vert \bar{r}\right\vert }}%
-2\tanh^{-1}\frac{\bar{m}_{B}}{\sqrt{\left\vert \bar{r}\right\vert }}%
+\tanh^{-1}\frac{1+\bar{m}_{B}-\bar{b}}{\sqrt{\left\vert \bar{r}\right\vert }%
}\right)  &\text{ if }\bar{r}>0\\
\frac{\bar{b}\bar{m}_{B}}{\left(  1-\bar{b}^{2}\right)  \bar{r}}-\frac{\bar
{b}}{\bar{r}\sqrt{\left\vert \bar{r}\right\vert }}\left(  \tan^{-1}%
\frac{1+\bar{m}_{B}+\bar{b}}{\sqrt{\left\vert \bar{r}\right\vert }}-2\tan
^{-1}\frac{\bar{m}_{B}}{\sqrt{\left\vert \bar{r}\right\vert }}+\tan^{-1}%
\frac{1+\bar{m}_{B}-\bar{b}}{\sqrt{\left\vert \bar{r}\right\vert }}\right)
&\text{ if }\bar{r}<0,
\end{array}
\right.  \label{p_ph implicit}%
\end{equation}
\end{widetext}
which changes smoothly across $\bar{r}=0$ at which $\bar{p}_{ph}%
=\alpha^{\left(  1\right)  }\left(  1+\bar{m}_{B}\right)  ^{2}2\bar{b}/\left(
3\bar{m}_{B}^{3}\right)$, and where $\bar{b}$ and $\bar{r}$ are functions of $p_{ph}$ given by
\begin{align}
\bar{b} &  =2\bar{m}_{B}\left(  \bar{p}-\bar{p}_{ph}\right) ,\quad 0<\bar{b}<1,\\
\bar{r} &  =\bar{m}_{B}^{2}+4\bar{m}_{B}^{2}\left(  \bar{p}-\bar{p}%
_{ph}\right)  ^{2}-1.
\end{align}

We now consider the HFB description encoded in $z_{k}$ and correlation
functions $\rho_{kk^{\prime}}$ and $\kappa_{kk^{\prime}}$. We solve for them
using the above MF solution as the initial guess in a self-consistent loop
which iteratively updates $z_{k}$, $\rho_{kk^{\prime}}$, and $\kappa_{kk^{\prime}}$
by solving Eqs.\ (\ref{CD equation}) and (\ref{AB equation}), in conjunction
with Eqs.\ (\ref{rho and kappa 1}) and (\ref{p_ph 1}), until values of a prescribed accuracy are reached.

\begin{figure}
[ptb]
\begin{center}
\includegraphics[width=3.3in]{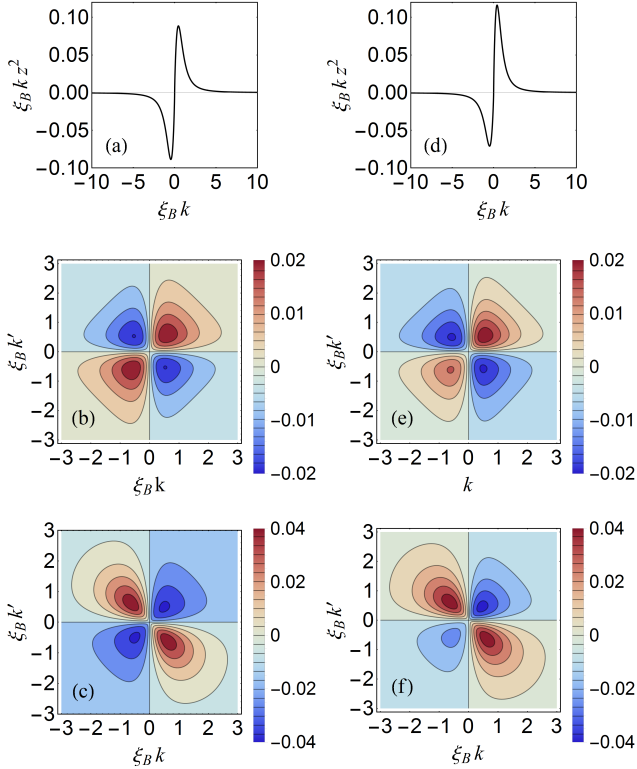}
\caption{(Color online) The phonon momentum density $kz_{k}^{2}$, the density correlation
$\rho_{k,k^{\prime}}$, and the pair correlation $\kappa_{k,k^{\prime}}$
characterizing a many-phonon system.  The first column is for the polaron ground state ($p=0$) and the second column is for a polaron state at finite momentum $p\xi_B = 1.5$.  Both columns have $m_B / m_I = 1$ and $\alpha^{(1)} = 2$.}
\label{Fig:phonon property}%
\end{center}
\end{figure}

The first column in Fig.\ \ref{Fig:phonon property}
displays an example with $p=0$, $m_{B}/m_{I}=1$, and $\alpha^{(1)}=2$. Interesting details emerge
from the 2D contour plots of $\rho_{kk^{\prime}}$ and $\kappa_{kk^{\prime}}$
in Figs.\ 1(b) and 1(c).  First, the vertical and horizontal lines at $k=0$ and
$k^{\prime}=0$ are the zero contour lines along which the correlation functions
vanish or are ``transparent."  This ``transparency" occurs because the effective phonon
interaction in momentum space is given by $kk^{\prime}/m_{I}$ [Eq.\
(\ref{four Boson interaction})] and thus vanishes when $k=0$ or $k^{\prime}%
=0$. Second, the $k=0$ and $k^{\prime}=0$ lines divide each contour plot
into two regions, one with $kk^{\prime}>0$ (the first and third quadrants) and
the other with $kk^{\prime}<0$ (the second and fourth quadrants). Each
correlation is seen to have opposite signs in these two regions. Third,
correlations develop peaks near but not at the origin, along the positive
diagonal ($k=k^{\prime}$) and negative diagonal ($k=-k^{\prime}$), but
decrease rapidly towards zero as momentum increases. This can be explained
as follows. As $k$ and $k^{\prime}$ increase, phonons interact more
strongly, but are tuned farther away from resonance, owing to an increase in
the effective single-phonon energy, $\mathcal{C}_{i}$ [Eq.\ (\ref{H_i})] in the
diagonal elements of matrix $A$ in Eq.\ (\ref{A and B}). In the limit of large
$k$ and $k^{\prime}$, being tuned away from resonance dominates and $\rho
_{kk^{\prime}}$ and $\kappa_{kk^{\prime}}$ become diminishingly small. The
peaks at intermediate momenta are the outcome of the competition between these
two opposing factors. 

The second column in Fig.\ \ref{Fig:phonon property} is the same as the
first column except $p=1.5\xi_{B}^{-1}$, e.g.\ a polaronic system prepared
adiabatically from one in which the impurity has a momentum $p=1.5\xi_{B}%
^{-1}$.  In contrast to the $p=0$ case, where $p_{ph}=0$ and all diagrams
[Figs.\ \ref{Fig:phonon property}(a), \ref{Fig:phonon property}(b), and \ref{Fig:phonon property}(c)] are symmetric, nonzero $p$ leads to nonzero $p_{ph}$ and an
asymmetry develops:\ the $k>0$ peak has a larger magnitude than the $k<0$
peak for $kz_{k}^{2}$ in Fig.\ \ref{Fig:phonon property}(d), with similar scenarios for the peaks along the diagonal
elements of the correlation functions, $\rho_{kk}$ and $\kappa_{kk}$ in Figs.\ \ref{Fig:phonon property}(e) and \ref{Fig:phonon property}(f). This is consistent
with the expectation that for nonzero $p$, a moving impurity drags a phonon
cloud with it, leading to nonzero phonon momentum $p_{ph}$.  However, nonzero
$p$ does not affect the symmetry of correlations between opposite momenta,
$\rho_{k,-k}$ and $\kappa_{k,-k}$, as can be seen in Figs.\ 1(e) and 1(f). The reason
is that $\rho$ and $\kappa$ are symmetric matrices and therefore $\rho_{k,-k}$ and $\kappa_{k,-k}$ must be even functions of $k$, independent of $p$.

\begin{figure}
[ptb]
\begin{center}
\includegraphics[width=3.4in]{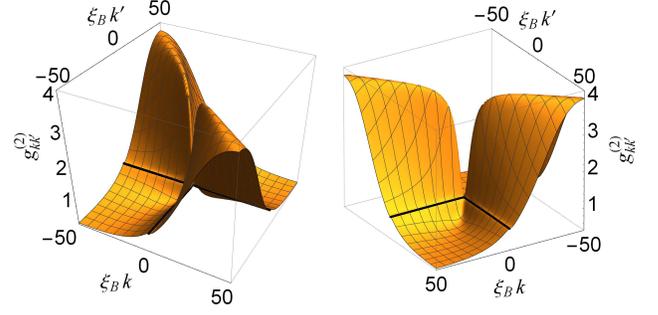}
\caption{(Color online) Two perspectives for the second order correlation function $g_{k,k^{\prime}}^{(2)}$ as a
function of $k$ and $k^{\prime}$ for the example in the first column of Fig.\ \ref{Fig:phonon property}.}
\label{Fig:g^2}
\end{center}
\end{figure}

In order to better understand the phonon cloud, such as the statistical character
of the quantum fluctuations, we follow \cite{shchadilova14arXiv:1410.5691} and
examine%
\begin{equation}
g_{kk^{\prime}}^{\left(  2\right)  }=\frac{\left\langle \hat{b}_{k}^{\dag}%
\hat{b}_{k^{\prime}}^{\dag}\hat{b}_{k^{\prime}}\hat{b}_{k}\right\rangle
}{\left\langle \hat{b}_{k}^{\dag}\hat{b}_{k}\right\rangle \left\langle \hat
{b}_{k^{\prime}}^{\dag}\hat{b}_{k^{\prime}}\right\rangle },
\end{equation}
which is the multi-mode generalization of the single-mode second-order
correlation, $g_{kk}^{(  2)  }$, popular in the study of quantum
optics \cite{walls08Book}, where $\langle \hat{b}_{k}^{\dag}\hat
{b}_{k^{\prime}}\rangle =z_{k}z_{k^{\prime}}+\rho_{kk^{\prime}}$ and
$\langle \hat{b}_{k}^{\dag}\hat{b}_{k^{\prime}}^{\dag}\hat{b}_{k^{\prime
}}\hat{b}_{k}\rangle =(  z_{k}z_{k^{\prime}}+\kappa_{kk^{\prime}%
})  ^{2}+z_{k}^{2}\rho_{k^{\prime}k^{\prime}}+z_{k^{\prime}}^{2}%
\rho_{kk}+2z_{k}z_{k^{\prime}}\rho_{kk^{\prime}}+\rho_{kk^{\prime}}^{2}%
+\rho_{kk}\rho_{k^{\prime}k^{\prime}}$, which are valid when quantities are real.  In Fig.\
\ref{Fig:g^2},
 the thick black lines passing through the origin indicate the $g_{kk^{\prime}}^{(2)  }=1$ plane (not shown), which is the value of $g_{kk^{\prime}}^{(2)  }$ if phonons are prepared in a MF coherent state.  The
region $kk^{\prime}<0$ exhibits phonon bunching, $g_{kk^{\prime}}^{(
2)  }>1$, while the region $kk^{\prime}>0$ exhibits phonon anti-bunching,
$g_{kk^{\prime}}^{(  2)  }<1$.  In particular, $g_{k,-k}^{(
2)  }$ decreases from 1 and saturates at a value less than 1 while
$g_{k,k}^{(  2)  }$ increases from 1 and saturates at a value
larger than $1$, a phenomenon first observed in an analogous 3D model
\cite{shchadilova14arXiv:1410.5691} and is believed to be accessible by noise
correlation analysis in time-of-flight experiments
\cite{ehud04PhysRevA.70.013603,simon05Nature.434.481,rom06Nature.434.481}.  A
qualitative explanation may be that in the region $kk^{\prime}<0$, the phonon
interaction is attractive and thus tends to cause phonons to cluster,
leading to phonon bunching, while in the region $kk^{\prime}>0$, the phonon
interaction is repulsive and thus tends to cause phonons to spread, leading to
phonon anti-bunching.

\subsection{Polaron Energy}

Having discussed the variables parameterizing the polaron, we now investigate
the polaron energy for a system with total momentum $p$. The
polaron energy was given in Eq.\ (\ref{<H>}), which may be simplified, with the
help of Eqs.\ (\ref{p_ph 1}) and (\ref{CD equation}), to
\begin{align}
E_{p} &  =\frac{p^{2}}{2m_{I}}+\frac{1}{\sqrt{\mathcal{V}}}\sum_{i}g_{_{i}%
}z_{i}-\frac{p_{ph}^{2}}{2m_{I}}\nonumber\\
&\qquad +  \sum_{i}\mathcal{C}_{i}\rho_{ii}+\frac{1}{2m_{I}}\sum_{i,j}\left(
k_{i}k_{j}\right)  \left(  \kappa_{ij}^{2}+\rho_{ij}^{2}\right)  ,\label{E2}%
\end{align}
which is valid in equilibrium where all variables are real.

As in the previous subsection, we begin with the MF limit where the trial
state is chosen as a product of coherent states parameterized by only $z_{k}$.
In this limit, the polaron energy (\ref{E2}) may be evaluated analytically and
gives (where $\bar{E}_{p}\equiv E_{p}/[\hbar^{2}/(  m_{B}\xi_{B}^{2})  ]$)
\begin{align}
\bar{E}_{p}  &  =\frac{\bar{m}_{B}}{2}\bar{p}^{2}-\frac{\bar{m}_{B}}{2}\bar
{p}_{ph}^{2}-\frac{\alpha^{\left(  1\right)  }}{2}\frac{\bar{m}_{B}^{2}}%
{\sqrt{\left\vert \bar{r}\right\vert }}\left(  1+\frac{1}{\bar{m}_{B}}\right)
^{2}\nonumber\\
&\quad  \times\left\{
\begin{array}{ll}
\coth^{-1}\frac{1+\bar{m}_{B}+\bar{b}}{\sqrt{\left\vert \bar{r}\right\vert }%
}+\coth^{-1}\frac{1+\bar{m}_{B}-\bar{b}}{\sqrt{\left\vert \bar{r}\right\vert
}} & \text{ if }\bar{r}>0\\
\cot^{-1}\frac{1+\bar{m}_{B}+\bar{b}}{\sqrt{\left\vert \bar{r}\right\vert }%
}+\cot^{-1}\frac{1+\bar{m}_{B}-\bar{b}}{\sqrt{\left\vert \bar{r}\right\vert }%
} &\text{ if }\bar{r}<0,
\end{array}
\right.   \label{Ep MF}%
\end{align}
which changes smoothly across $\bar{r}=0$ at which $\bar{E}_{p} = \bar{m}_B[
\bar{p}^{2}-\bar{p}_{ph}^{2}-\alpha^{\left(  1\right)  }\left(  1+\bar{m}%
_{B}^{-1}\right)  ^{2}]  /2$.  The polaron energy $\bar{E}_{p}$
depends on the total momentum $p$.  However, it has been long established
\cite{spohn86JPhysicsA.19.533} that the ground state, where the polaron energy
is lowest, occurs at $p=0$.  This is a general statement, and is thus true
for both the HFB and MF descriptions.  For the MF description, the ground
state polaron energy is then obtained from Eq.\ (\ref{Ep MF}) by setting
$p=0$:
\begin{equation}
\bar{E}_{0}=-\alpha^{\left(  1\right)  }\frac{\left(  1+\bar{m}_{B}\right)
^{2}}{\sqrt{\left\vert \bar{m}_{B}^{2}-1\right\vert }}\left\{
\begin{array}{ll}
\coth^{-1}\frac{1+\bar{m}_{B}}{\sqrt{\left\vert \bar{m}_{B}^{2}-1\right\vert
}} & \text{ if }\bar{m}_{B}>1\\
\cot^{-1}\frac{1+\bar{m}_{B}}{\sqrt{\left\vert \bar{m}_{B}^{2}-1\right\vert }%
} & \text{ if }\bar{m}_{B}<1,
\end{array}
\right.   \label{E0 MF}%
\end{equation}
and $\bar{E}_{0}=-2\alpha^{(1)}$ when $\bar{m}_{B}=1$. 

We benchmark our HFB model by comparing its prediction for the ground state polaron energy with the predictions from MF theory (\ref{E0 MF}) and Feynman's path integral formalism, which was regarded as a superior
all coupling approximation \cite{tempere09PhysRevB.80.184504}. Feynman's
method amounts to applying the Feynman-Jensen inequality on a
variational action describing two (classical) particles coupled via a harmonic
force, where one is the impurity and the other is a fictitious particle.  
Steps involved in integrating out the degrees of freedom for the fictitious
particle leading to an effective variational action for the impurity are
highlighted in Appendix A.

The first column in Fig.\ \ref{Fig:Polaron Energy}
\begin{figure}
[ptb]
\begin{center}
\includegraphics[
width=3.2093in
]%
{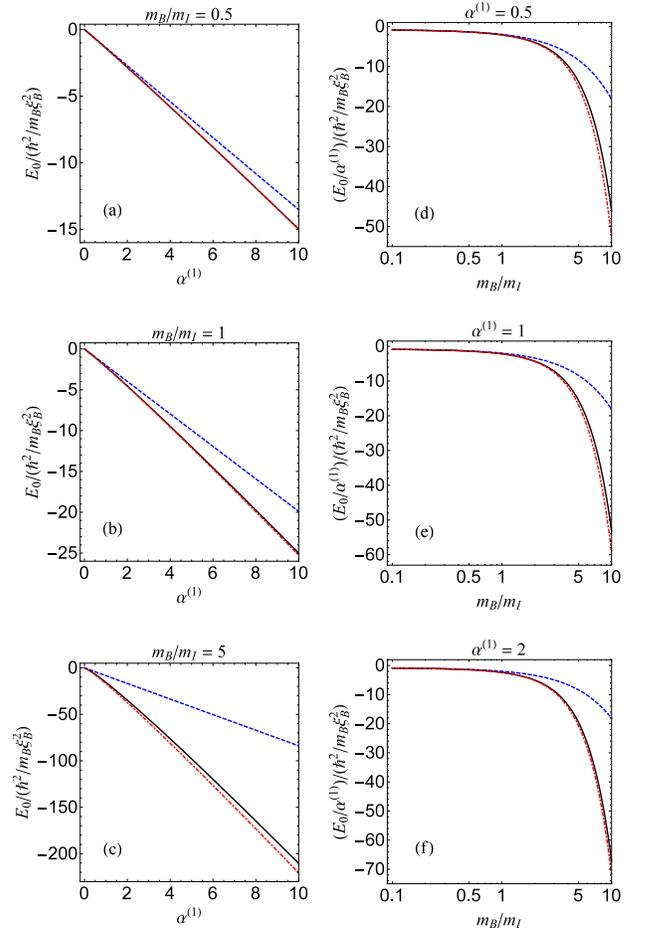}%
\caption{(Color online) The first column displays the ground state polaron energy $E_{0}$ in units of $\hbar^{2}/\left(  m_{B}\xi_{B}^{2}\right)  $ as a function of the
dimensionless polaronic coupling constant $\alpha^{\left(  1\right)  }$ when
(a) $m_{B}/m_{I}=0.5$, (b) $1$, and (c) $5$. The second column shows the
polaron energy divided by $\alpha^{\left(  1\right)  },$ $E_{0}/\alpha
^{\left(  1\right)  }$, in units of $\hbar^{2}/\left(  m_{B}\xi_{B}%
^{2}\right)  $ as a function of the boson-impurity mass ratio, $m_{B}/m_{I}$,
when (d) $\alpha^{\left(  1\right)  }=0.5$, (e) $1$, and (f) $5$.  In each plot the solid black curve is our HFB result, the dashed blue curve is the MF result, and the dash-dotted red curve is Feynman's path integral result.}%
\label{Fig:Polaron Energy}%
\end{center}
\end{figure}displays the ground state
polaron energy, $\bar{E}_{0}$, as a function of the coupling constant
$\alpha^{(  1)  }$ for various boson-impurity mass ratios, $\bar
{m}_{B}=m_{B}/m_{I}$.  The dashed blue curves are obtained from the MF theory [Eq.\
(\ref{E0 MF})], the solid black curves are obtained from our HFB theory, and the dash-dotted red
curves from Feynman's path integral formalism.  The MF variational ansatz
for finite $\bar{m}_{B}$ is motivated by the observation that the MF theory
becomes exact in the limit of heavy impurity $\bar{m}_{B}\rightarrow0$, where
$\hat{H}_{int}$ in Eq.\ (\ref{H after LLP}) is negligible and the shifting
operation (\ref{shifting}) with $z_{k}=-(g_{k}/\omega_{k})/\sqrt{\mathcal{V}}$
alone can diagonalize Eq.\ (\ref{H after LLP}).  Indeed, results from all
three approaches, although not shown, would be plotted virtually atop one
another for roughly $\bar{m}_{B}<0.2$.  As the impurity becomes increasingly
less massive, i.e.\ $\bar{m}_{B}$ increases, Figs.\ \ref{Fig:Polaron Energy}(a),
\ref{Fig:Polaron Energy}(b), and \ref{Fig:Polaron Energy}(c) illustrate that the MF results become increasingly larger than
Feynman's, in sharp contrast to the HFB results which match nicely with
Feynman's, demonstrating that correlations, which are excluded from the MF
theory, are an important part of the ground polaron state

The second column in Fig.\ \ref{Fig:Polaron Energy} displays $\bar{E}%
_{0}/\alpha^{(  1)  }$ as a function of the mass ratio, $\bar
{m}_{B}$, for various values of the coupling constant $\alpha^{(
1)  }$. Equation (\ref{E0 MF}) tells us the MF $\bar{E}_{0}$ is
proportional to $\alpha^{(  1)  }$ and thus $\bar{E}_{0}%
/\alpha^{(  1)  }$ is independent of $\alpha^{(  1)  }$,
as illustrated by identical dashed curves in the second column. In the limit
of heavy impurity mass, Eq.\ (\ref{E0 MF}) asymptotes to
\begin{equation}
\bar{E}_{0}/\alpha^{\left(  1\right)  }\approx-\frac{\pi}{4}+\frac{1}{2}%
\bar{m}_{B}-\frac{5\pi}{8}\bar{m}_{B}^{2}+\cdots,\label{E0 heavy}%
\end{equation}
where, as explained above, the MF result becomes an exact solution.  As can
be seen from the second column, the HFB and Feynman results agree very well
with the MF results in this limit.  In the limit of light impurity, Eq.\
(\ref{E0 MF}) asymptotes to
\begin{equation}
\frac{\bar{E}_{0}}{\alpha^{\left(  1\right)  }}
\approx-\ln\left(  2\bar{m}_{B}\right)
\left(  \frac{\bar{m}_{B}}{2}+1\right)  +\frac{1-6\ln\left(  2\bar{m}%
_{B}\right)  }{8\bar{m}_{B}}+\cdots.\label{E0 light}%
\end{equation}
In this case we do not expect the MF result to be accurate and we find again
that the HFB and Feynman results disagree strongly with the MF results but
agree well with each other, indicating as before that neglecting quantum
fluctuations in the light impurity limit can lead to significant errors. The HFB and Feynman energies are seen to decrease rapidly with decreasing
impurity mass (increasing $\bar{m}_{B}$), while the MF energy changes slowly
due to the existence of a logarithmic function in the leading term in Eq.\
(\ref{E0 light}).

\subsection{Effective Polaron Mass}

Finally, we turn our attention to the effective polaron mass $m_{I}^{\ast}$
defined by
\begin{equation}
m_{I}^{\ast}=\left(  \left.  \frac{\partial^{2}E_{p}}{\partial p^{2}%
}\right\vert _{p=0}\right)  ^{-1}, \label{m*}%
\end{equation}
which follows from expansion of the polaron energy through second order in the
total momentum $p$, $E_{p}\approx E_{0}+p^{2}/2m_{I}^{\ast}$, where $E_{0}$ is
the ground state polaron energy studied in Fig.\ \ref{Fig:Polaron Energy}.
$m_{I}^{\ast}$ emerges naturally from Landau's concept of a mobile
polaron, in which an impurity drags with it a cloud of nearby background
particles, leading to an effective mass $m_{I}^{\ast}$ heavier than its bare
mass $m_{I}$. This picture together with the conservation of momentum means
the impurity momentum $p_{I}$ equals the total momentum minus the momentum of
the phonon cloud $p_{ph}$: $p_{I}=p-p_{ph}$, leading to the formula
\cite{shashi14PhysRevA.89.053617}
\begin{equation}
\frac{1}{m_{I}^{\ast}}=\frac{1}{m_{I}}-\frac{1}{m_{I}}\lim_{p\rightarrow
0}\frac{p_{ph}}{p},\text{ } \label{1/m_I}%
\end{equation}
which is consistent with Eq.\ (\ref{m*}). 

\begin{figure}
[ptb]
\begin{center}
\includegraphics[
width=3.3in
]%
{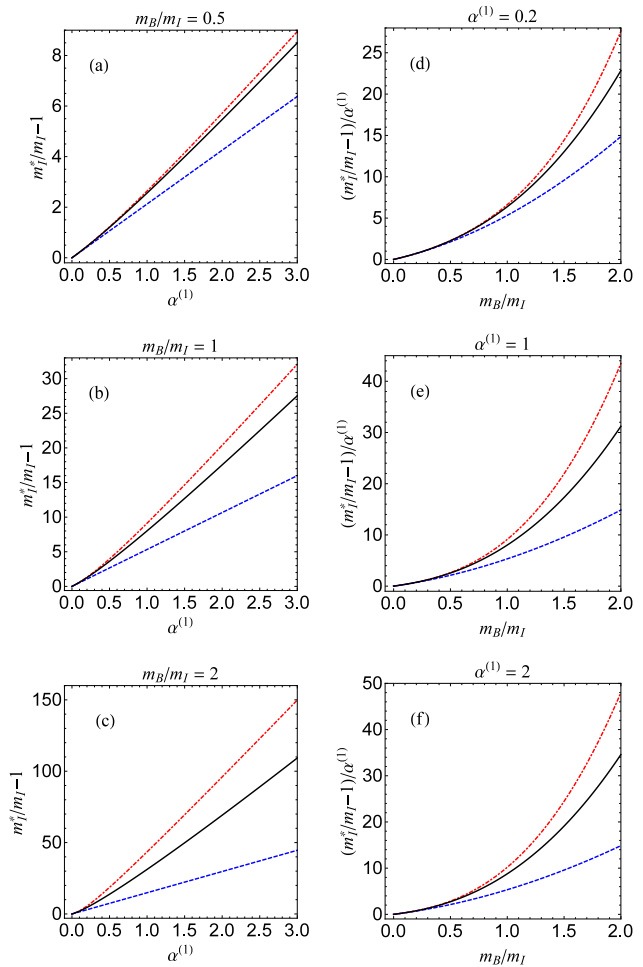}%
\caption{(Color online) A comparison of effective polaron masses according to MF theory (dashed blue curves), our HFB theory (solid black curves), and Feynman's variational method
(dash-dotted red curves).  In the first column, $m_{I}^{\ast}/m_{I} -1$ is plotted as a function of $\alpha^{(1)}$ for (a) $m_{B}=0.5$, (b) $1.0$, and (c)
$2.0$.  In the second column, $(m_I^*/m_I - 1)/\alpha^{(1)}$ is plotted as a function of $m_{B}/m_{I}$ for
(d) $\alpha^{(1)}=0.5$, (e) $1.0$, and (f)
$2.0$.}%
\label{Fig:polaron mass}%
\end{center}
\end{figure}

Figure \ref{Fig:polaron mass}
displays the effective polaron mass $m_{I}%
^{\ast}$. We show $\bar{m}_{I}^{\ast}$ as a function of
$\alpha^{\left(  1\right)  }$ for various values of $\bar{m}_{B}$ in the first column and as a function of $\bar{m}_{B}$ for various values of
$\alpha^{\left(  1\right)  }$ in the
second column. In both columns the solid black curves are from our
HFB method, and the dashed blue curves are from the MF theory, which, as in the
previous subsection, can be computed analytically:
\begin{align}
&\bar{m}_{I}^{\ast}   =1+\frac{2\alpha^{\left(  1\right)  }\bar{m}_{B}%
^{2}\left(  1+\bar{m}_{B}\right)  }{\bar{m}_{B}-1}+\frac{4\alpha^{\left(
1\right)  }\bar{m}_{B}\left(  1+\bar{m}_{B}\right)  }{\left(  \bar{m}%
_{B}-1\right)  \sqrt{\left\vert \bar{m}_{B}^{2}-1\right\vert }}
\nonumber\\
& \times \left\{
\begin{array}{ll}
\tanh^{-1}\frac{1+\bar{m}_{B}}{\sqrt{\left\vert \bar{m}_{B}^{2}-1\right\vert
}}-\tanh^{-1}\frac{\bar{m}_{B}}{\sqrt{\left\vert \bar{m}_{B}^{2}-1\right\vert
}}&\text{ if }\bar{m}_{B}>1,\\
-\tan^{-1}\frac{1+\bar{m}_{B}}{\sqrt{\left\vert \bar{m}_{B}^{2}-1\right\vert
}}+\tan^{-1}\frac{\bar{m}_{B}}{\sqrt{\left\vert \bar{m}_{B}^{2}-1\right\vert
}} & \text{ if }\bar{m}_{B}<1,
\end{array}
\right.
\end{align}
and $\bar{m}_{I}^{\ast}=1+\frac{16}{3}\alpha^{(  1)  }$ if
$\bar{m}_{B}=1$.  Figure \ref{Fig:polaron mass} also includes the effective
mass obtained from Feynman's approach using Eq.\ (\ref{m* Feynman}) in Appendix
A as the dash-dotted red curves. Figure \ref{Fig:polaron mass} demonstrates
that the HFB theory consistently gives a heavier effective mass than the MF
theory and that it can be significantly heavier for small $\bar{m}_{B}$ or
large $\alpha^{(  1)  }$.  The effective mass using Feynman's
method, while consistently heavier than in both methods, is much closer to our
HFB result, once again demonstrating the nonclassical nature of the phonon
cloud inside of which phonons are highly correlated.  A difference between
Feynman's and our HFB masses is expected since Feynman's approach cannot compute
the polaron energy at finite $p$ and hence defines the effective mass
differently from Eq.\ (\ref{m*}). 

We conclude this subsection by noting that in the heavy impurity limit, Eq.\
(\ref{m* Feynman}) is found numerically to agree with the MF result
\begin{equation}
\bar{m}_{I}^{\ast}\approx1+\alpha^{\left(  1\right)  }\pi\bar{m}_{B}%
+\alpha^{\left(  1\right)  }\left(  2\pi-4\right)  \bar{m}_{B}^{2}%
+\cdots,\label{m_I MF}%
\end{equation}
while the variational mass $M$ is found to depart significantly from the above
MF result.  Thus, in Feynman's method, $M$ does not agree with the effective polaron mass formula in Eq.\ (\ref{m* Feynman}) and we must use Eq.\ (\ref{m* Feynman}) to compute the effective polaron mass.

\section{Conclusion}

We considered the Fr\"ohlich model in a moving frame defined by the LLP transformation, where the original impurity-phonon system is transformed to an interacting many-phonon system free of impurities. This LLP model distinguishes itself with the four-boson interaction term in Eq.\ (\ref{four Boson interaction}) where an interaction between two phonons with momentum $\mathbf{k}$ and $\mathbf{k}'$ does not involve any
momentum exchange and is facilitated by a
``potential" that depends on $\mathbf{k}\cdot\mathbf{k}'$. In the
spirit of generalized HFB theory, we formulated a field theoretical description of the LLP model where phonons are subject to this unique phonon-phonon interaction.  As an application, we applied our theory to Bose polarons in quasi-1D cold atom mixtures and investigated polaron properties such as energy and mass by solving the HFB equations self-consistently and the HFB equations in the MF limit analytically.

We found in the regime of relatively light impurity and strong coupling, our HFB results were significantly closer to those from Feynman's method than predictions from MF theory.  The agreement between our HFB approach and Feynman's method on the polaron energy was particularly impressive.  We found in the strongly interacting region that the polaron ground state contains highly correlated phonon pairs. In any many-body system at (or close to) zero temperature, the exact nature of the ground state depends crucially on how particles interact with each other.  We attributed  the existence of both repulsive (in the region
$kk^{\prime}>0$) and attractive (in the region $kk^{\prime}<0$) phonon-phonon
interactions to the rich structure exhibited in various correlation functions, and to bunching and anti-bunching statistics exhibited in the second-order correlation function.

We expect the 3D polaron to behave differently than our 1D polaron since their densities of states differ.  Nevertheless, it is worth pointing out that for the 3D polaron, as the polaron coupling constant increases the ground state energy first rises above the impurity-condensate
interaction energy and then decreases below it, while in our 1D case the
ground state energy is always below it and decreases monotonically with increasing
$\alpha^{\left(  1\right)  }$.  This difference may be traced to the fact
that the 3D case suffers from an ultraviolet divergence, a
complication that does not occur in our 1D model.  As a result, the 1D system
has allowed us to focus our attention on our main purpose:\ gaining clean
insight into the role the effective phonon-phonon interaction and quantum
fluctuations play in polaronic states. 

Finally, we comment that the recent upsurge of interest in Bose polarons has
been largely spurred by the prospect that the rich toolbox and the flexibility
of cold atom systems may allow polaron theories to be tested, to great
precision, in cold atom experiments. However, many observables, which occur as
correlation functions involving various field operators, are inaccessible to
Feynman's method.  Our HFB theory, however, can cast these observables
into forms which are, at least in principle, amenable to numerical analysis.
As a concrete example, in Appendix B we express, in terms of the variational
parameters of polarons, a time-dependent overlap function that lies at the
heart of radio-frequency (rf) spectroscopy, which has emerged as a
powerful tool in the study of cold atom physics in general and polaron physics
in particular.

\section*{Acknowledgments}

B.K.\ is grateful to ITAMP and the Harvard-Smithsonian Center for Astrophysics for their hospitality during the beginning stages of this work.  H.Y.L.\ is supported in part by the U.S. National Science Foundation under Grant No.\ PHY 11-25915.

\appendix

\section{Feynman's Variational Approach}

In this appendix, we outline Feynman's variational approach to the polaron
\cite{feynman55PhysRev.97.660}, beginning with the partition function,
$Z=$Tr$(  e^{-\beta\hat{H}^{\prime}})  $, where $\hat{H}^{\prime}$
is given by Eq.\ (\ref{H-Frohlich}) and $k_{B}\beta$ equals the inverse
temperature.  Tracing out the bosonic degrees of freedom gives rise
to an effective action for the impurity%
\begin{align}
S &  =\int_{0}^{\beta}d\tau\frac{1}{2}m_{I}\mathbf{\dot{r}}^{2}\left(
\tau\right)  -\frac{1}{2\mathcal{V}}\sum_{\mathbf{k}}g_{\mathbf{k}}^{2}%
\nonumber\\
&\quad \times  \int_{0}^{\beta}\int_{0}^{\beta}d\tau d\sigma\mathcal{G}\left(
\omega_{\mathbf{k}},\left\vert \tau-\sigma\right\vert \right)  e^{i\mathbf{k}%
\cdot\left[  \mathbf{r}\left(  \tau\right)  -\mathbf{r}\left(  \sigma\right)
\right]  },\label{S action}%
\end{align}
where
\begin{equation}
\mathcal{G}\left(  x,u\right)  =\cosh\left[  x\left(  u-\beta/2\right)
\right]  /\sinh\left(  \beta x/2\right)  ,
\end{equation}
with $\omega_{\mathbf{k}}$ and $g_{\mathbf{k}}$ given in Eqs.\
(\ref{w_k}) and (\ref{g_k}), respectively. 

To compute the ground state free energy, $F$, of the Fr\"{o}hlich Hamiltonian
$\hat{H}^{\prime}$ in Eq.\ (\ref{H-Frohlich}) [hence the action in Eq.\
(\ref{S action})], Feynman introduces a novel variational approach based on
the Feynman-Jensen inequality,
\begin{equation}
F\leq F_{var}-\left\langle S-S_{var}\right\rangle _{S_{var}},\label{F Fvar}%
\end{equation}
where $F_{var}$ is the free energy of a variational system and the average in
$\langle S-S_{var}\rangle _{S_{var}}$ is performed with respect to
the variational system's ground state. Minimizing the right-hand side of Eq.\
(\ref{F Fvar}) gives the strongest bound on $F$, which we take to be our
estimate of $F$.  As a variational system, Feynman uses the impurity
interacting with a fictitious particle via a harmonic potential.  Integrating
out the fictitious particle yields the variational action for the impurity:
\begin{align}
S_{var} &  =\int_{0}^{\beta}d\tau\frac{1}{2}m_{I}\mathbf{\dot{r}}^{2}\left(
\tau\right)  +\frac{MW^{3}}{8}\nonumber\\
& \times  \int_{0}^{\beta}\int_{0}^{\beta}d\tau d\sigma\mathcal{G}\left(
W,\left\vert \tau-\sigma\right\vert \right)  \left[  \mathbf{r}\left(
\tau\right)  -\mathbf{r}\left(  \sigma\right)  \right]  ^{2},\label{S var}%
\end{align}
where $M$, the mass of the fictitious particle, and $W$, the frequency of the
harmonic potential, are the variational parameters. A straightforward but
lengthy calculation gives for the inequality (\ref{F Fvar})
\cite{tempere09PhysRevB.80.184504,casteels12PhysRevA.86.043614}
\begin{align}
F &  \leq\frac{d}{\beta}\left[  \ln\sinh\left(  \frac{\beta\Omega}{2}\right)
-\ln\sinh\left(  \frac{\beta\Omega}{2\sqrt{1+M/m_{I}}}\right)  \right]
\nonumber\\
&  -\frac{d}{2\beta}\ln\frac{m_{I}+M}{m_{I}}-\frac{d}{2\beta}\frac{M}{m_{I}%
+M}\left[  \frac{\beta\Omega}{2}\coth\frac{\beta\Omega}{2}-1\right]
\nonumber\\
&  -\sum_{\mathbf{k}}g_{\mathbf{k}}^{2}\int_{0}^{\beta/2}du\, \mathcal{G}\left(
\omega_{\mathbf{k}},u\right)  \mathcal{M}_{M,\Omega}\left(  \mathbf{k}%
,u\right)  ,\label{F<}%
\end{align}
where $d$ is the dimension, $W$ is replaced in favor of the more convenient
$\Omega=W\sqrt{1+(  M/m_{I})  }$, and
\begin{align}
&  \mathcal{M}_{M,\Omega}\left(  \mathbf{k},u\right)  \nonumber\\
&  =\exp\left\{  -\frac{k^{2}}{2\left(  m_{I}+M\right)  }\left[  u-\frac
{u^{2}}{\beta}+\frac{M}{m_{I}\Omega}\left(  \cdots\right)  \right]  \right\}
,
\end{align}
with $\left(  \cdots\right)  $ being defined as
\begin{equation}
\left(  \cdots\right)  =\frac{\cosh\left(  \beta\Omega/2\right)  -\cosh\left[
\Omega\left(  \beta/2-u\right)  \right]  }{\sinh\left(  \beta\Omega/2\right)
}.
\end{equation}

As discussed in the main body of the paper, our interest lies in the zero
temperature limit, which simplifies Eq.\ (\ref{F<}) to%
\begin{align}
E_{0} &  \leq\frac{\Omega d}{2}-\frac{\Omega d}{2\sqrt{1+M/m_{I}}}%
-\frac{M\Omega d}{4\left(  m_{I}+M\right)  }\nonumber\\
&  -\frac{1}{\mathcal{V}}\sum_{\mathbf{k}}g_{\mathbf{k}}^{2}\int_{0}^{\infty
}due^{-u\omega_{\mathbf{k}}}\lim_{\beta\rightarrow\infty}\mathcal{M}%
_{M,\Omega}\left(  \mathbf{k},u\right)  ,\label{E0 Feynman}%
\end{align}
where we have changed the free energy $F$ to the polaron ground state energy $E_{0}$ and
\begin{align}
&  \lim_{\beta\rightarrow\infty}\mathcal{M}_{M,\Omega}\left(  \mathbf{k}%
,u\right)  \nonumber\\
&  =\exp\left\{  -\frac{k^{2}}{2\left(  m_{I}+M\right)  }\left[
u+\frac{M\left(  1-e^{-u\Omega}\right)  }{m_{I}\Omega}\right]  \right\}  .
\end{align}
At zero temperature, Feynman \cite{feynman55PhysRev.97.660} derived from the
asymptotic form of the partition function a formula for the effective polaron
mass,
\begin{align}
m_{I}^{\ast} &  =m_{I}+\frac{1}{d\mathcal{V}}\sum_{\mathbf{k}}k^{2}%
g_{\mathbf{k}}^{2}\nonumber\\
&\qquad \times  \int_{0}^{\infty}due^{-u\omega_{\mathbf{k}}}u^{2}\lim_{\beta\rightarrow
\infty}\mathcal{M}_{M,\Omega}\left(  \mathbf{k},u\right)  .\label{m* Feynman}%
\end{align}

The positive $M$ and $\Omega$ that minimize the right hand side of Eq.\
(\ref{E0 Feynman}) for $d=1$ are fed back into the left hand side of Eq.\
(\ref{E0 Feynman}) to give the polaron energy $E_{0}$ which we display in Fig.\
\ref{Fig:Polaron Energy}. These same values of $M$ and $\Omega$ are then
used in Eq.\ (\ref{m* Feynman}) for $d=1$ to compute the effective mass which
we display in Fig.\ \ref{Fig:polaron mass}.

\section{RF Spectrum}

In this appendix, we establish a framework for calculating the impurity rf spectra
\cite{chin04Science305.20082004,schirotzek09PhysRevLett.102.230402,feld11Nature.480.75}
in our HFB model. In rf spectroscopy, an rf field of amplitude $F_{L}$ and
frequency $\omega_{L}$ is applied to promote the impurity from an initial
state $\vert g\rangle $ to a final state $\vert e\rangle$, which are internal states, e.g.\ hyperfine states. The two states differ by
energy $\omega_{eg}$ and the process is described by the Hamiltonian
$\tilde{H}_{rf}=(  F_{L}e^{-i\omega_{L}t}\vert e\rangle
\langle g\vert + \text{h.c.})  /2$.  The total Hamiltonian
$\tilde{H}$ for this two-state impurity-BEC system is
\begin{equation}
\tilde{H}=\omega_{eg}\left\vert e\right\rangle \left\langle e\right\vert
+\hat{H}_{gg}\left\vert g\right\rangle \left\langle g\right\vert +\hat{H}%
_{ee}\left\vert e\right\rangle \left\langle e\right\vert +\tilde{H}%
_{rf},\label{H total}%
\end{equation}
where $\hat{H}_{gg}$ and $\hat{H}_{ee}$ describe impurity-phonon subsystems
containing $\vert g\rangle $-type and $\vert e\rangle $-type
impurities, respectively. The polarization of such a system is expected to
oscillate periodically at rf frequency $\omega_{L}$, and the rf (or probe
absorption) spectrum is then expected to be proportional to the following
expectation value:
\begin{equation}
\operatorname{Re}\left[  i\left\langle \left\vert g\right\rangle \left\langle
e\right\vert \right\rangle e^{i\omega_{L}t}/F_{L}\right].\label{rf probe spectrum}
\end{equation}

Let $\vert i_{g}\rangle $ be the ground state of $\hat{H}_{gg}$
with energy $E_{ig}$ and $\vert f_{e}\rangle $ be any eigenstate of
$\hat{H}_{ee}$ with energy $E_{fe}$. Let $\vert i_{g},g\rangle $
and $\vert f_{e},e\rangle $ be the initial and final
(impurity-phonon) states of the total system described by $\tilde{H}$ in Eq.\
(\ref{H total}). Evaluating Eq.\ (\ref{rf probe spectrum}), within the
framework of linear response theory, yields, straightforwardly, the rf
spectrum (Fermi's golden rule),
\begin{equation}
I_{p}\left(  \omega\right)  =\sum_{f_{e}}\left\vert \left\langle
f_{e},e\right\vert e\rangle \langle g\left\vert i_{g},g\right\rangle \right\vert
^{2}\delta\left[  \omega-\left(  E_{fe}-E_{ig}\right)  \right]  ,\label{I_p}%
\end{equation}
where the subscript $p$ on the left-hand side (which we suppress on the
right-hand side to reduce clutter) represents the total momentum [first
introduced in Eq.\ (\ref{H after LLP})] and the frequency $\omega\equiv
\omega_{L}-\omega_{eg}$ is measured relative to the $\vert e\rangle
\longleftrightarrow\vert g\rangle $ atomic transition.  A standard
manipulation transforms Eq.\ (\ref{I_p}) into an integral
\cite{knap12PhysRevX.2.041020}
\begin{equation}
I_{p}\left(  \omega\right)  =\operatorname{Re}\frac{1}{\pi}\int_{0}^{\infty
}dte^{i\omega t}\mathcal{A}_{p}\left(  t\right)  ,\label{I(p,w) 1}%
\end{equation}
involving an overlap function in the time domain defined as
\begin{equation}
\mathcal{A}_{p}\left(  t\right)  =e^{iE_{ig}t}\left\langle i_{g}\right\vert
e^{-i\hat{H}_{ee}t}\left\vert i_{g}\right\rangle .\label{Ap(t)}%
\end{equation}

In the (direct) rf measurement, the rf field excites impurities in state $\vert
g\rangle $, which interact with the BEC via $s$-wave scattering, to
state $\vert e\rangle $, where they do not interact with the BEC. In
this case, $\hat{H}_{gg}$ is the interacting phonon Hamiltonian in Eq.\
(\ref{H after LLP}), $\hat{H}_{ee}$ is the free phonon Hamiltonian $\hat{H}_{ee}=\sum_{\mathbf{k}}\omega_{\mathbf{k}}\hat{b}_{\mathbf{k}}^{\dag
}\hat{b}_{\mathbf{k}}$ \cite{shashi14PhysRevA.89.053617}, $E_{ig}$ is the polaron
energy $E_{p}$ in Eq.\ (\ref{<H>}), and $\vert i_{g}\rangle $ is the
polaron state $\vert \phi\rangle $ introduced in Sec.\ II.

To facilitate the evaluation of Eq.\ (\ref{Ap(t)}), we express the polaron
state $\vert \phi\rangle $ in terms of the phonon vacuum
$\vert 0\rangle $ \cite{blaizot96QuantumTheoryBook}
\begin{equation}
\left\vert \phi\right\rangle =\frac{\exp\frac{1}{2}\sum_{\mathbf{kk}^{\prime}%
}\hat{c}_{\mathbf{k}}^{\dag}\left(  G_{\mathbf{kk}^{\prime}}^{\ast}%
\equiv\left(  U^{\ast-1}V^{\ast}\right)  _{\mathbf{kk}^{\prime}}\right)
\hat{c}_{\mathbf{k}^{\prime}}^{\dag}}{\left[  \det\left(  U^{\dag}U\right)
\right]  ^{1/4}}\left\vert 0\right\rangle ,
\end{equation}
and at the same time organize $\exp(  -i\hat{H}_{ee}t)  $ into an
antinormal ordered form \cite{louisell90Book}
\begin{align}
&\exp\left(  -i\hat{H}_{ee}t\right)     =
{\displaystyle\prod_{\mathbf{k}}}
\sum_{m=0}^{\infty}\frac{\left(  1-\exp\left(  i\omega_{\mathbf{k}}t\right)
\right)  ^{m}}{m!}\nonumber\\
&\qquad \times  \exp\left(  i\omega_{\mathbf{k}}t\right)  \left(  z_{\mathbf{k}}+\hat
{c}_{\mathbf{k}}\right)  ^{m}\left(  z_{\mathbf{k}}^{\ast}+\hat{c}%
_{\mathbf{k}}^{\dag}\right)  ^{m}.
\end{align}

Let $\vert c_{\mathbf{k}}\rangle $ be the coherent state of
$\hat{c}_{\mathbf{k}}$, i.e.\ $\hat{c}_{\mathbf{k}}\vert c_{\mathbf{k}%
}\rangle =c_{\mathbf{k}}\vert c_{\mathbf{k}}\rangle ,$
normalized according to $\langle c_{\mathbf{k}}|0\rangle =1$.  In
the coherent state space defined by the completeness relation
\begin{equation}
{\displaystyle\prod_{\mathbf{k}}}
\int\frac{dc_{\mathbf{k}}^{\ast}dc_{\mathbf{k}}}{2\pi i}e^{-c_{\mathbf{k}%
}c_{\mathbf{k}}^{\ast}}\left\vert c_{\mathbf{k}}\right\rangle \left\langle
c_{\mathbf{k}}\right\vert =I,
\end{equation}
we can cast the overlap function, $\mathcal{A}_{p}\left(  t\right)
=e^{iE_{p}t}\left\langle \phi\right\vert e^{-i\hat{H}_{ee}t}\left\vert
\phi\right\rangle ,$ into a Gaussian integral
\begin{align}
\mathcal{A}_{p}\left(  t\right)   &  =\exp\sum_{\mathbf{k}}\left[
i\omega_{\mathbf{k}}t+\left(  1-e^{i\omega_{\mathbf{k}}t}\right)  \left\vert
z_{\mathbf{k}}\right\vert ^{2}\right]  \nonumber\\
&  \qquad\times\frac{e^{iE_{p}t}}{\sqrt{\det\left(  U^{\dag}U\right)  }}
{\displaystyle\prod_{\mathbf{k}}}
\int\frac{dc_{\mathbf{k}}^{\ast}dc_{\mathbf{k}}}{2\pi i}e^{h},\label{Ap(t) 2}%
\end{align}
where%
\begin{equation}
h=-c^{\ast}Kc+\frac{1}{2}cGc+\frac{1}{2}c^{\ast}G^{\ast}c^{\ast}+xc+y^{\ast
}c^{\ast},
\end{equation}
or explicitly
\begin{align}
\mathcal{A}_{p}\left(  t\right)   &  =e^{iE_{p}t}\frac{\exp\sum_{\mathbf{k}%
}\left[  i\omega_{\mathbf{k}}t+\left(  1-e^{i\omega_{\mathbf{k}}t}\right)
\left\vert z_{\mathbf{k}}\right\vert ^{2}\right]  }{\sqrt{\det\left(  U^{\dag
}U\right)  }}\nonumber\\
& \times  \frac{\exp\left\{  \frac{1}{2}\left(  x,y^{\ast}\right)  \left(
\begin{array}
[c]{cc}%
K & G^{\ast}\\
G & K
\end{array}
\right)  ^{-1}\left(
\begin{array}
[c]{c}%
y^{\ast}\\
x
\end{array}
\right)  \right\}  }{\sqrt{\det\left(
\begin{array}
[c]{cc}%
K & G^{\ast}\\
G & K
\end{array}
\right)  }},\label{A_p final}%
\end{align}
where $K$ is a diagonal matrix defined as%
\begin{equation}
K_{ij}=\exp\left(  i\omega_{\mathbf{k}_{i}}t\right)  \delta_{i,j},
\end{equation}
and $x$ and $y$ are vectors defined as
\begin{align}
x_{i} &  =\left(  1-\exp\left(  i\omega_{\mathbf{k}_{i}}t\right)  \right)
z_{\mathbf{k}_{i}}^{\ast},\\
y_{i} &  =\left(  1-\exp\left(  -i\omega_{\mathbf{k}_{i}}t\right)  \right)
z_{\mathbf{k}_{i}}^{\ast}.
\end{align}

Equation (\ref{A_p final}) is the main result of this appendix, which
expresses the overlap function (\ref{Ap(t)}) and hence the rf spectrum
(\ref{I(p,w) 1}) in terms of the variables that parametrize the HFB
variational polaron state.  This may be extended, using the time-dependent
HFB variational principle, to inverse rf spectroscopy
\cite{kohstall12Nature.485.615,shashi14PhysRevA.89.053617}, where the rf field
transfers impurities in state $\vert g\rangle $, which do not
interact with the BEC, to state $\vert e\rangle $, where they do interact
with the BEC.  We leave this as a possible future research project.


\end{document}